\documentclass[12pt,a4paper]{article}
\usepackage[utf8]{inputenc}
\usepackage[T1]{fontenc}
\usepackage{lmodern}
\usepackage{amsmath,amssymb}
\usepackage{graphicx}
\usepackage[linesnumbered,ruled,vlined]{algorithm2e}
\usepackage{caption}
\usepackage{subcaption}
\usepackage{hyperref}
\usepackage{geometry}
\usepackage{setspace}
\usepackage{algorithmic}
\usepackage{algorithm2e}
\usepackage{pgfplots}
\usepackage{fancyhdr}

\fancypagestyle{firstpage}{
  \fancyhf{} 
  \fancyfoot[C]{\scriptsize
    © 2025 IEEE. Personal use of this material is permitted. For all other uses, in any current or future media, permission must be obtained from IEEE.\\
    This is the accepted version of the paper to appear in the 2025 IEEE Global Blockchain Conference (IEEE GBC 2025).\\
    The final published version will be available at: [ ]
  }

}

\title{Fully Decentralised Consensus for Extreme-scale Blockchain}

\author{
  Siamak Abdi\thanks{Free University of Bolzano, Italy. Email: sabdi@unibz.it} \and
  Giuseppe Di Fatta\thanks{Free University of Bolzano, Italy. Email: giuseppe.difatta@unibz.it} \and
  Atta Badii\thanks{University of Reading, United Kingdom. Email: atta.badii@reading.ac.uk} \and
  Giancarlo Fortino\thanks{University of Calabria, Italy. Email: giancarlo.fortino@unical.it}
}

\date{July 2025}

\begin{document}
\maketitle
\thispagestyle{firstpage}
\begin{abstract}
Blockchain is a decentralised, immutable ledger technology that has been widely adopted in many sectors for various applications such as cryptocurrencies, smart contracts and supply chain management. 
Distributed consensus is a fundamental component of blockchain, which is required to ensure trust, security, and integrity of the data stored and the transactions processed in the blockchain. 
Various consensus algorithms have been developed, each affected from certain issues such as node failures, high resource consumption, collusion, etc. 
This work introduces a fully decentralised consensus protocol, Blockchain Epidemic Consensus Protocol (BECP), suitable for very large and extreme-scale blockchain systems. The proposed approach leverages the benefits of epidemic protocols, such as no reliance on a fixed set of validators or leaders, probabilistic guarantees of convergence, efficient use of network resources, and tolerance to node and network failures. 
A comparative experimental analysis has been carried out with traditional protocols including PAXOS, RAFT, and Practical Byzantine Fault Tolerance (PBFT), as well as a relatively more recent protocol such as Avalanche, which is specifically designed for very large-scale systems. The results illustrate how BECP outperforms them in terms of throughput, scalability and consensus latency. BECP achieves an average of 1.196 times higher throughput in terms of consensus on items and 4.775 times better average consensus latency. Furthermore, BECP significantly reduces the number of messages compared to Avalanche. 
These results demonstrate the effectiveness and efficiency of fully decentralised consensus for blockchain technology based on epidemic protocols.
\end{abstract}

\section{Introduction}
Blockchain is a Distributed Ledger Technology (DLT) that stores transactions or any items in the form of sequential blocks. 
A distributed ledger ensures that data is replicated across multiple nodes for transparency and security. Each block contains a hash of the previous block, creating an immutable chain of blocks. Any blockchain network relies on a consensus mechanism that ensures agreement on the states of certain data or items among distributed participants \cite{ferdous2020blockchain}. For example, in cryptocurrencies like Bitcoin and Ethereum, blockchain technology is utilized to validate and record transactions securely and transparently across a decentralised network of nodes. Consensus in blockchain generally refers to the process of agreeing on the validity of transactions and ensuring the integrity and security of the blockchain. Achieving consensus among nodes in a distributed network is one of the critical challenges in blockchain systems \cite{singh2022survey}, \cite{hussein2023evolution}. 

While there are several consensus algorithms available, such as Paxos \cite{lamport2001paxos}, Raft \cite{ongaro2014search}, Practical Byzantine Fault Tolerance (PBFT) \cite{lamport2019byzantine}, Proof-of-x (e.g. Proof-of-Work (PoW), Proof-of-Stake (PoS)), and Avalanche \cite{rocket2018snowflake}, each has limitations that make them less ideal for certain use cases. For example, traditional protocols Paxos, Raft, and PBFT are effective in closed or permissioned networks but unsuitable for open or public networks. These protocols rely on the existence of a centralized entity or leader, and assume a fully connected communication network, where nodes know and trust each other and can communicate directly with one another (one-to-all and all-to-one). Having a centralized leader makes the system vulnerable to node failures as it is a bottleneck and a single point of failure in the system, and requires all nodes to trust the leader \cite{blasa2011symmetric}. Furthermore, it is unreasonable to expect that each node can communicate directly with all other nodes in a large-scale dynamic network. Reaching global consensus in a fully decentralised fashion within a distributed system without full connectivity is an interesting and important challenge \cite{cason2021gossip}.

Consensus mechanisms have shifted from centralized models to distributed and decentralised ones \cite{singh2022survey} in order to overcome these limitations and meet the need for scalability, fault tolerance, availability, data replication, and reliability of data. The removal of a centralized authority or leader from the consensus process also reduces system vulnerability and mitigates the impact of malicious nodes. 

Proof-of-x-based algorithms are well-known decentralised consensus mechanisms, but they have their own problems. PoW consumes a lot of resources and energy and may exhibit slowness and inefficiency. Accordingly, they are not suitable for some application domains such as Internet-of-Thing (IoT) due to their high resource consumption \cite{singh2022survey}. In PoS, another widespread consensus mechanism, validator nodes hold and stake tokens for the privilege of earning transaction fees.
PoS can be susceptible to problems related to centralization and to attacks by groups of wealthy nodes. This makes this mechanism more vulnerable to corruption or collusion \cite{hussein2023evolution}. 

Avalanche \cite{rocket2018snowflake} is a relatively recent decentralised consensus protocol, which works with the combination of a node sampling method and an epidemic diffusion approach to provide a consensus mechanism in a distributed network. Sampling-based algorithms like Avalanche face challenges in achieving global consensus or convergence and often exhibit significant network overhead in large networks. For example, in Avalanche nodes are required to inquire their neighbours to get informed of their votes repeatedly. This approach causes two contrasting problems in the protocol as it is difficult to set the sample size parameter to an appropriate number of neighbours: small values can significantly increase the latency of consensus in large networks and adopting large values may overwhelm the network with messages.

To address the limitations of existing consensus algorithms, we introduce Blockchain Epidemic Consensus Protocol (BECP), a consensus approach based on epidemic protocols. BECP leverages a randomised communication and computation methodology. The novelty of this work is that unlike the proof-of-based algorithms, which use Gossiping solely for informing participating nodes about new blocks (information dissemination), our method extends its utilization for achieving consensus by means of fully decentralised data aggregation. This approach offers outstanding scalability and robust fault tolerance, and inherits the fast (logarithmic) convergence property of epidemic communication with acceptable cost. Epidemic approaches are leaderless, so there is no bottleneck in the system, and they provide a strong probabilistic guarantee that every participating node will eventually receive the required information and converge to consensus \cite{ayiad2016agreement}, \cite{blasa2011symmetric}.

The rest of the paper is outlined as follows. Section \ref{sec:related_work} reviews related works of existing consensus protocols in blockchain networks. Section \ref{sec:BECP_method} describes the details of the BECP consensus method. Section \ref{sec:implementation} delves into the implementation of the studied protocols, along with comprehensive network configuration details.
Section \ref{sec:performance_evaluation} defines the performance evaluation methodology used for evaluating the performance of BECP. Section \ref{sec:discussion} discusses the results and findings. Finally, conclusions and future work are concluded in Section \ref{sec:conclusion}.
\section{Related Work}
\label{sec:related_work}
A consensus algorithm is defined as a protocol or mechanism that is used to achieve agreement among the nodes on a particular item in a distributed network \cite{hussein2023evolution}. An effective and inclusive consensus algorithm involves all participants in making decisions based on conflicts within blockchain networks \cite{singh2022survey}. To date, numerous consensus algorithms have been proposed; the choice of underlying consensus mechanism depends on the type of blockchain and how it is used \cite{singh2022survey}. Generally, we can classify consensus mechanisms into three broad categories: direct communication-based, proof-of-x-based, and epidemic-based.

Direct communication-based (traditional) consensus algorithms were designed to address the challenge of achieving consensus among a group of nodes in a closed or permissioned network where nodes know each other and any new node is verified before joining the network. In such a network, nodes may experience failures or exhibit malicious behaviour, which makes consensus difficult to achieve. 

Paxos is a consensus algorithm that was first introduced by Lamport in 1998 \cite{lamport2001paxos}. It is a leaderless and three-phase algorithm in which all nodes have the same role and communicate with each other to reach a consensus on a value. Raft is another consensus algorithm that was designed to improve upon the problems of Paxos. It was introduced by Ongaro and Ousterhout in 2014 \cite{ongaro2014search}. PBFT algorithm is one of this class of consensus algorithms that can tolerate byzantine (malicious) behaviours in a distributed system which constitutes $3f + 1$ nodes with $f$ malicious nodes. In other words, a consensus can be reached as long as no less than $2f + 1$ or 66\% of non-byzantine nodes are functioning normally \cite{lamport2019byzantine}.

Paxos, Raft, and PBFT are highly influential algorithms in distributed networks and have served as the basis for many other consensus algorithms. Several other algorithms such as MultiPaxos \cite{chand2016formal}, Fast Paxos \cite{lamport2006fast}, Byzantine Paxos \cite{lamport2011byzantizing}, Delegated Byzantine Fault Tolerance (DBFT), and Federated Byzantine Agreement (FBA) \cite{singh2022survey} based on these foundational algorithms have been developed and are currently utilized in blockchain systems to ensure agreement among nodes or participants \cite{lashkari2021comprehensive}, \cite{ferdous2020blockchain}, \cite{singh2022survey}. 

Blockchain networks utilize proof-of-x-based consensus algorithms to ensure that all nodes reach a consensus on the state of the network, including the order of transactions and the balance of accounts. Unlike traditional algorithms that have a centralized authority, these algorithms are decentralised. The algorithms reach an agreement by proofing mechanisms and solve inconsistency or forks—instances where two or more conflicting blocks emerge simultaneously—with approaches like selecting the longest chain (Nakamoto protocol) or choosing the heaviest chain in the GHOST protocol \cite{singh2022survey}, \cite{ferdous2020blockchain}, \cite{rocket2018snowflake}. The two most popular consensus algorithms in the Proof-based category used in blockchain are PoW and PoS \cite{jakobsson1999proofs}, \cite{thin2018formal}.
An emerging category of consensus algorithms used in distributed ledger systems such as blockchain networks is epidemic-based consensus algorithms. The 'snow' family consensus algorithms \cite{rocket2018snowflake} belong to this class which provide a decentralised consensus mechanism with epidemic communication and a sampling method. These algorithms are performed by sampling or inquiring from a small set of participants (neighbours) and reaching an agreement based on predefined measurements. Avalanche is the latest member of the snow family that is utilized in blockchain networks. 

On the general problem of agreement and consensus in distributed systems, several works have adopted an epidemic approach. 

In \cite{ayiad2016agreement}, the authors propose a novel protocol named Phase Transition Protocol (PTP) to achieve global consensus on the convergence of a distributed information dissemination process. The protocol uses an epidemic approach to exchange information between random nodes and employs a local computation to achieve distributed consensus.

In \cite{ayiad2017agreement}, the authors introduce the Epidemic Consensus Protocol (ECP), which is employed to achieve consensus in distributed data aggregation. ECP achieves consensus by iteratively exchanging information between random nodes until all nodes converge to a common state with explicit local detection of global convergence.

The existing works introduced on consensus mechanisms have some disadvantages that make them less suitable for application in blockchain networks. Traditional algorithms such as PAXOS, RAFT, and PBFT are deterministic, reaching consensus in predefined phases of message passing. However, they are practical only in closed, private, or permissioned blockchain networks. Additionally, the existence of a centralized entity, like a leader, makes them inconsistent with the primary objectives of blockchain networks. Proof-based mechanisms address the centralization problem of traditional algorithms but face challenges such as high computational requirements (PoW) or vulnerability to collusion (PoS). 

The recent epidemic-based consensus algorithm, Avalanche, offers a more affordable and reliable decentralised mechanism, but it suffers from a significant volume of sent messages due to its sampling approach. To address these gaps, we introduce a fully decentralised consensus protocol leveraging the advantages of epidemic communication. It employs lightweight local calculations instead of a sampling approach, providing a well-suited consensus protocol for blockchain networks.
\section{The Consensus Protocol}
\label{sec:BECP_method}
The proposed method is inspired by the works \cite{ayiad2016agreement}, \cite{ayiad2017agreement}, \cite{kempe2003gossip}, and \cite{blasa2011symmetric} to devise a fully decentralised consensus protocol that is suitable for the distributed computation in blockchain networks with no reliance on a fixed set of validators or leaders, which also provides solid probabilistic guarantees of convergence and has efficient use of network resources.\hspace{0.5em}In the remainder of this section, the details of the proposed algorithm are described. 
\begin{figure}[!htbp]
    \centering
    \begin{subfigure}[b]{0.7\textwidth}
        \centering
        \includegraphics[clip, trim=1mm 1mm 1mm 1mm, width=\textwidth]{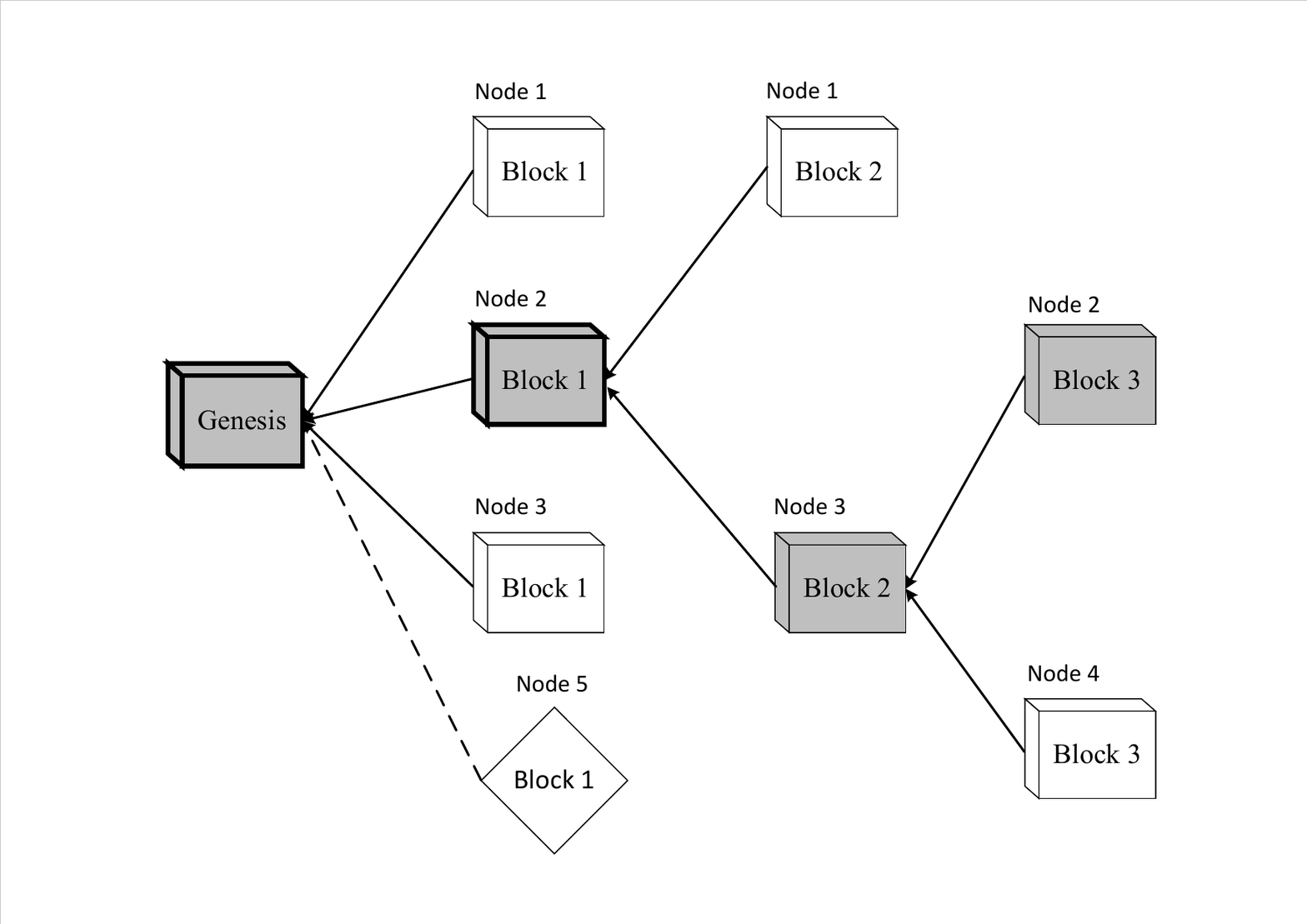}
        \caption{Case I}
        \label{fig:resolution_revised}
    \end{subfigure}
    
    \vspace{1em}  
    
    \begin{subfigure}[b]{0.7\textwidth}
        \centering
        \includegraphics[clip, trim=1mm 1mm 1mm 1mm, width=\textwidth]{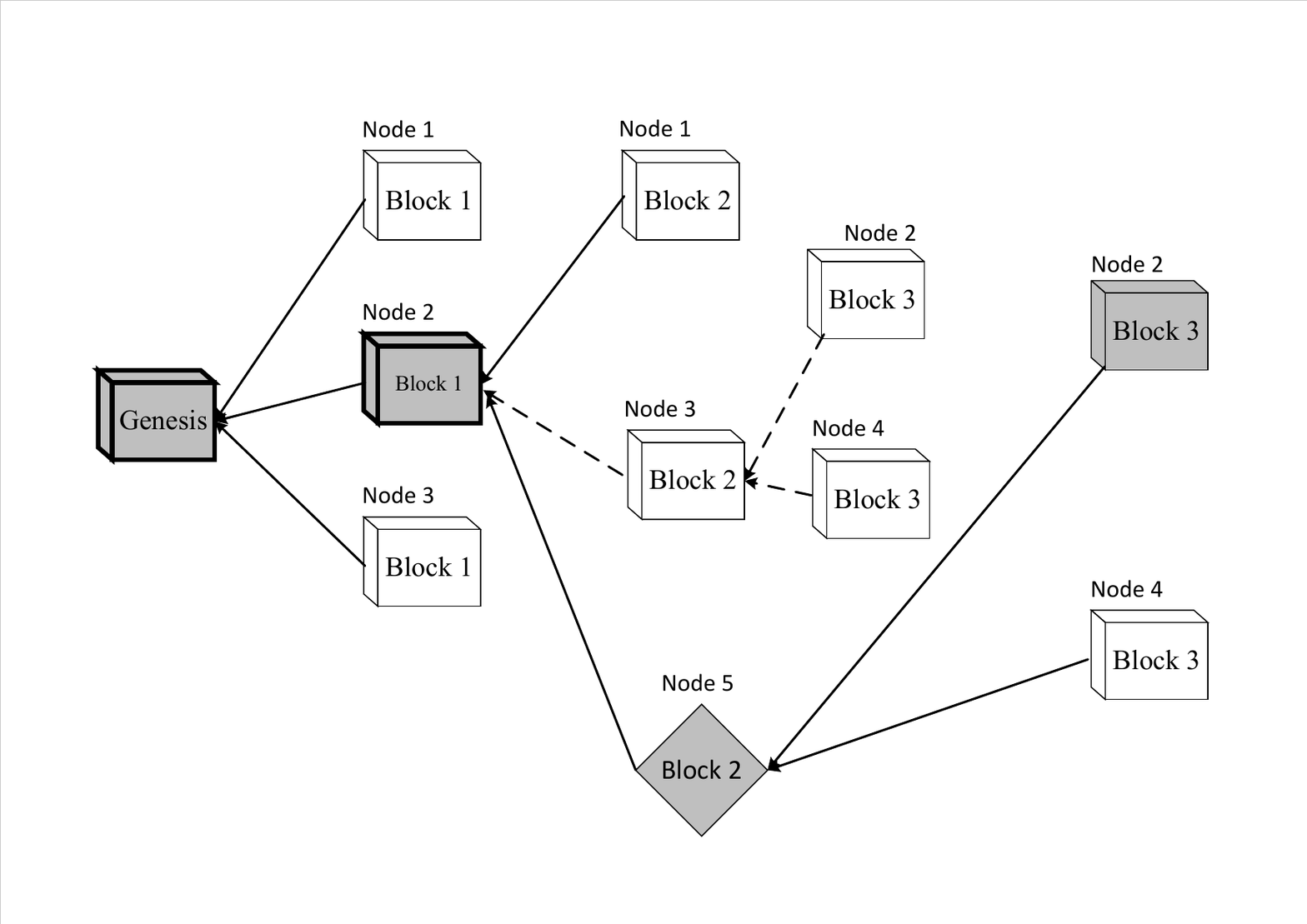}
        \caption{Case II}
        \label{fig_resolution_backward}
    \end{subfigure}
    
    \caption{Illustration of Duplicate Block Resolution (Case I) and Backward Scenario (Case II): White blocks represent candidate blocks that dropped, pale blocks denote the currently preferred blocks, the diamond block indicates the newly received block, and blocks with highlighted edges signify confirmed blocks. The text over the blocks indicates the node that created each block.}
    \label{fig:resolution}
\end{figure}
\label{sec:BECP}
The Blockchain Epidemic Consensus Protocol (BECP) is a fully decentralised epidemic consensus approach applied to blockchain technology. BECP consists of three interrelated protocols: the System Size Estimation Protocol (SSEP), the Node Cache Protocol (NCP), and the Phase Transition Protocol (PTP). SSEP continuously monitors the system size, offering the function \textit{getSystemSize()}, which accurately estimates the number of participating nodes in a blockchain system in real time. NCP provides a scalable membership sampling function, \textit{getRandomNode()}, by randomly selecting nodes for epidemic communication purposes without complete knowledge of the system, which can be dynamic. Finally, the PTP operates as a decentralised consensus algorithm, making use of the other two protocols and resolving issues related to duplicates and guarantees unique IDs in blocks.

The original PTP protocol in \cite{ayiad2016agreement} has been modified and adapted to work with blockchain. Utilizing the original PTP with the current algorithm does not apply to blockchain networks directly because of the nature and structure of blockchain; in a blockchain, each item or block on which consensus occurs should refer to its parent (previous block). After generating a new block, its content and reference cannot be changed since they are encrypted. 

The problem in the original PTP protocol arises since items are generated and accepted without having references. However, in one approach, if participants generate their blocks by referring to their last generated blocks, there is a possibility of breaking the chain of references when a block is accepted among multiple candidate blocks with similar IDs. An alternate solution to tackle this issue is that nodes should wait for the confirmation of the last block and then generate their new blocks by referring to that block. Once a candidate block is chosen as a confirmed block among other candidate blocks, others will be eliminated from the local caches since they are not valid anymore. This approach, although it guarantees a chain of blocks with correct references, is highly inefficient since nodes must wait for the confirmation of blocks. This decreases the throughput of the systems. 

To address this issue more efficiently, we propose a new consensus procedure in algorithm \ref{algo:revised-PTP} for accepting and resolving inconsistent or duplicate blocks, blocks with similar IDs or similar parents. The consensus procedure in BECP consists of two parts: first, duplicate blocks are resolved in the local caches; then, for the selected blocks, the estimation process is performed. Once the state of a block changes to commit, identified through the estimation process, a consensus is reached on the block. The second part, where consensus is reached on the block, is more time-consuming. As depicted in the algorithm, we define a new variable named $B_{pref}$, which represents the current preferred block, and impose nodes to generate new blocks by referring to that block. In this way, nodes can generate blocks without the need for the confirmation of the last block.

The current preferred block $B_{pref}$ is the block that is chosen as the possible and potential confirmed block among other candidate blocks in the nodes' local block cache. Once a node receives a block that is selected as the new preferred block, it will accept the new block, and discard the current preferred block along with its descendant blocks. (Algorithm \ref{algo:backward}). Afterwards, the node will generate a new block by referring to the new preferred block as needed. This approach besides guaranteeing a consistent chain of blocks with correct references, increases throughput since nodes do not wait for the confirmation of the last block.

Algorithm \ref{algo:revised-PTP} is a revised version of the Resolve Duplicate Block ID Procedure, which is used to determine $B_{pref}$ and insert new blocks into local block caches. First, nodes compare the received block with all existing blocks in the local cache to check for duplicates (line 2)—specifically, if there is an unconfirmed block with an identical ID or similar parents. They must resolve this duplication by choosing either the received block $\tau'$ or the existing block $\tau$. The first if statement in line 3 of Algorithm \ref{algo:revised-PTP} checks if $\tau'$ is the same as $\tau$. In that case, the procedure drops $\tau'$ but uses its pairs ($vp$, $wp$, $va$, and $wa$) to update the corresponding pairs in $\tau$. Otherwise, the procedure checks if $\tau'$ is selected as $B_{pref}$ (line 5 and 6). If this is the case, the procedure executes the backward procedure, removes $\tau$, adds $\tau'$ into the local cache, and sets $\tau'$ as $B_{pref}$. $\tau'$ is chosen as a preferred block $B_{pref}$ based on the condition ($\tau'.t = \tau.t \text{ and } \tau'.o < \tau.o$) or ($\tau'.t < \tau.t$). In other words, a preferred block $B_{pref}$ is chosen based on the block’s generation time and the ID of its originators.

If there is no block with an identical ID to the received block $\tau'$, it is added to the local cache (line 14). This addition depends on verifying whether $\tau'$ is connected to the current $B_{pref}$ by checking that the creator node of $\tau'$'s parent is the same as the creator node of $B_{pref}$. Alternatively, this can be done by comparing the hash of block $B_{pref}$ with the hash of $\tau'$'s parent block. This ensures a consistent and ordered chain of blocks.

Figure \ref{fig:resolution} (case I) illustrates the process of resolving duplicate or inconsistent blocks. All newly generated blocks refer to the current preferred block (pale blocks), and other candidate blocks (white blocks) are dropped. There is always a preferred block in the local cache. However, in the figure, the diamond-shaped block with the height of $1$ is received after all other blocks. Since block 1 is confirmed (the block with highlighted edges), nodes drop the received diamond-shaped block. Figure \ref{fig:resolution} (case II) depicts a scenario where a backwards occurs: Before receiving the diamond-shaped block 2, there was a preferred block (block 2, created by node 3) and its BECPendants in the chain. When the diamond-shaped block is received, nodes execute the backward procedure, removing the previous preferred block (block 2, created by node 3) and its BECPendants, meaning they are not valid anymore, and setting the diamond-shaped block as the new preferred block. Subsequently, new blocks are connected to this new preferred block.

Furthermore, we offer a new code snipped (Algorithm \ref{algo:block_generation}) designed for the block generation process, which prevents nodes from double-spending and producing blocks with identical IDs and creators. In the algorithm, nodes verify that there are no existing blocks with identical IDs created by themselves, and the new block references the current preferred block. This precautionary step helps prevent potential occurrences of double-spending. Subsequently, upon successful generation of a new block, nodes will set the new block as their current preferred block $B_{pref}$. We assume that communication is reliable and that all nodes are benign.
\\

\begin{algorithm}[H]
  \caption{Backward Procedure}
  \label{algo:backward}
  \SetAlgoLined
  \KwData{Received Block $\tau'$}
  \KwResult{Discarding Invalid Blocks}
  \SetKwProg{Fn}{Function}{:}{}
  \Fn{\textsc{backward}(block local cache $C_b$, $\tau'$}{
    $\text{children} \leftarrow \text{$\tau'$.getChildren()}$\;
    \ForEach{$\text{child} \in \text{children}$}{
      \textit{\textsc{backward} ($\text{$C_b$, child}$)\;}
    }
    \text{remove $\tau'$ from the $C_b$}\;
  }
\end{algorithm}
\begin{algorithm}[H]
  \caption{Block Generation Procedure}
  \label{algo:block_generation}
  \KwResult{Generate a new block $B_i$}
  \SetKwProg{Fn}{Function}{:}{}
  \Fn{\textsc{generateNewBlock()}}{
    \If{$\text{block local cache $C_b$ does not contain a block with id}$
       \hspace{1em}$(B_{pref}.id + 1)$}{
      \text{ generate a new block $B_i$}
      \text{add $B_i$ to the children set of $B_{pref}$}
      \text{add $B_i$ into the block local cache $C_b$}\;
      \text{set $B_i$ to $B_{pref}$}
    }
  }
\end{algorithm}
\begin{table*}
  \small
  \renewcommand{\arraystretch}{1.3}
  \caption{Experiment Settings}
  \label{table:experiment_settings}
  \centering
  \begin{tabular}{|c|c|c|c|c|c|}
    \hline
    Parameters & BECP & Avalanche & PBFT & PAXOS & RAFT \\
    \hline
    D1 (seconds) & 0.1 & 0.1 & $-$ & $-$ & $-$ \\
    Network Latency & [0.01,0.3) & [0.01,0.3) & [0.01,0.3) & [0.01,0.3) & [0.01,0.3) \\
    Cycle Time (seconds) & 0.7 & 0.7 & $-$ & $-$ & 0.7 \\
    $K$ & $-$ & 10 & $-$ & $-$ & $-$ \\
    $\alpha$ & $-$ & 0.8 & $-$ & $-$ & $-$ \\
    $\beta_1$ & $-$ & 50 & $-$ & $-$ & $-$ \\
    $\beta_2$ & $-$ & 150 & $-$ & $-$ & $-$ \\
    $T_{\text{block}}$ (seconds) & 10 & 10 & 10 & 10 & 10 \\
    $P_{\text{block}}$ & 5\% & 5\% &  $-$ &  $-$ &  $-$ \\
    $\epsilon_1$ & 0.01 & $-$ & $-$ & $-$ & $-$ \\
    $N_{\text{cache}}$ & 50 & $-$ & $-$ & $-$ & $-$ \\
    $\psi$ & 3 & $-$ & $-$ & $-$ & $-$ \\
    Timeout Range & $-$ & $-$ & $-$ & $-$ & 1.0 to 1.2 \\
    \hline
  \end{tabular}
\end{table*}
\section{Implementation and Network Configuration}
\label{sec:implementation}
To implement and evaluate the BECP protocol alongside other studied blockchain consensus protocols in this work, a blockchain simulator is required. The Just Another Blockchain Simulator (JABS) \cite{yajam2023jabs} was chosen as the ideal block simulation environment due to its several advantageous features. It simplifies the simulation of blockchain networks and makes it easier to understand and work with. Additionally, it offers a high degree of customization, allowing us to adjust the simulation parameters to suit our project requirements. It also supports various network configurations, enabling us to explore different scenarios. The simulator is written in Java programming language, providing the advantage of fast execution and usability.

JABS operates as a discrete-event simulator, where events are characterized as messages received within a node. Specifically, events occur when a node receives a message from a peer. The simulation starts by executing initial events generated by the nodes. During this phase, each node generates a block with a timestamp and sends it to its peer. Furthermore, the nodes continue to generate new blocks at fixed intervals of time with a defined probability.

We implemented the Avalanche protocol with a blockchain framework rather than a DAG structure. Each newly generated block in this framework points to the block with the highest ID within the known set of nodes. This approach enables the nodes to form a chain of blocks, facilitating a fair comparison between systems.

We conducted simulations using the PAXOS, RAFT, PBFT, BECP, and Avalanche protocols on the JABS simulator, employing specific parameters. Our simulations comprised two main parts. In the first part, the simulations were run for $3600$ seconds (one hour) for PAXOS, RAFT, PBFT, and BECP, and other simulations were run for $600$ seconds ($10$ minutes) for BECP and Avalanche due to the complexity of Avalanche's running time. The simulations were run with seed values of $1$ to $5$, and a block generation interval of $10$ seconds. These simulations were conducted on a Linux server with 400GB of RAM. The settings for each experiment are detailed in the following.

All the protocols were simulated on a WAN network topology where message latency was modeled using a uniform probability function with parameters set to a minimum of $0.01$ seconds and a maximum of $0.3$ seconds. In the protocols BECP, Avalanche, and Raft, nodes are activated in periodic intervals, therefore we defined the parameter cycle time and set it to $0.7$ seconds. In the other two protocols, nodes are activated when they receive a message from the leader. Furthermore, due to the structure of BECP and Avalanche, nodes can propose their blocks simultaneously hence we define an initial interval time $D_1$ for nodes and set it to $0.1$ seconds. In Paxos, Raft, and PBFT only the leader node proposes the blocks, so there is no need to define this parameter. 

The parameters of BECP were configured as follows. The $\epsilon_1$ value, representing the estimation error threshold, was set at $0.01$, and the $\psi$, the minimum number of consecutive cycles threshold, was configured at $3$ cycles. The $N_{\text{cache}}$ representing the number of neighbours' IDs stored in the cache was equal to $50$. In Avalanche, sample size ($K$), quorum size threshold ($\alpha$), safe early commitment threshold ($\beta_1$), and consecutive counter threshold ($\beta_2$) were set to $10$, $0.8$, $50$, and $150$.

The block generation process differed between the deterministic protocols (PAXOS, RAFT, and PBFT), and the probabilistic protocols (BECP and Avalanche). In the traditional protocols, initially, a leader generates a new block and starts the consensus process. After reaching a consensus on the block, the leader node can generate another new block. Contrastingly, Avalanche and BECP nodes can propose their blocks within an interval without waiting for the confirmation of preceding blocks, which is aligned more closely with real-world applications. In our simulations, we set the parameters $T_{\text{block}}$ (block generation interval) and $P_{\text{block}}$ (block generation probability) to $10$ seconds and $5\%$ respectively for BECP and Avalanche. For the other protocols, we set $T_{\text{block}}$ to $10$ seconds.

In the simulator, blocks and transactions were sampled from a distribution based on Bitcoin block and transaction sizes. The message size was determined as the cumulative sum of block sizes in the local block cache. Table \ref{table:experiment_settings} provides a concise overview of the protocol settings.\\ 

\begin{algorithm}[H]
  \caption{Revised Resolve Duplicate Block ID Procedure}
  \label{algo:revised-PTP}
  \ForEach{$\tau' \in m.C$}{
    \If{$C$ contains $\tau$ where $\tau'.id = \tau.id$}{
      \If{$\tau$ has not been confirmed}{
        \If{$\tau'.t = \tau.t$ \textbf{and} $\tau'.o = \tau.o$}{
          $\tau \gets \langle \tau.id, \tau.o, \tau.t, \tau.vp + \tau'.vp, \tau.wp +
          \tau'.wp, \tau.va + \tau'.va, \tau.wa + \tau'.wa, \tau.\text{state} \rangle$
        }
        \ElseIf{$(\tau'.t = \tau.t \; \textbf{and} \; \tau'.o < \tau.o) \; \textbf{or} \; (\tau'.t < \tau.t)$}{
          \textsc{backward} (\text{local cache}, $\tau$)
          $\tau \gets \langle \tau'.id, \tau'.o, \tau'.t, \tau'.vp + 1, 
          \tau'.wp, \tau'.va, \tau'.wa, \tau'.\text{state} \rangle$ 
          \textbf{set $B_{pref}$ to $\tau'$}
        }
      }
    }
    \ElseIf{\text{creator node of parent of } $\tau'.p$ = \text{creator node of } $B_{pref}$}{
      $C \gets C \cup \{\langle \tau'.id, \tau'.o, \tau'.t, \tau'.vp + 1,
      \tau'.wp, \tau'.va, \tau'.wa, \tau'.\text{state} \rangle\}$
      \textbf{set $B_{pref}$ to $\tau'$}
    }
  }
\end{algorithm}
\section{Performance Evaluation}
\label{sec:performance_evaluation}
In this section, to assess the performance of BECP, we conduct a thorough performance comparison with existing protocols of PAXOS, RAFT, PBFT, and Avalanche. Our evaluation includes an in-depth interpretation of the results obtained from the experiments. In the following, we define and describe the measurements of Throughput, Scalability, Communication Overhead, and Consensus Latency which are important metrics for any blockchain network.  

\textbf{Throughput:} Throughput stands out as an important metric for blockchain networks, representing the number of items accepted by nodes within a given time frame. A higher value of Throughput represents the effectiveness of the associate consensus mechanism. An item is defined as a block in the protocols PAXOS, RAFT, PBFT, and BECP and a transaction in the protocol Avalanche based on their structures.

\textbf{Scalability:} Scalability is another crucial metric for blockchain networks. Scalability determines how well a protocol can perform in large-scale networks. Figure \ref{fig:throughput_1} and Figure \ref{fig:throughput_2} illustrate a comparison of throughput and scalability metrics among the protocols.
\begin{figure}
    \centering
    \begin{tikzpicture}
        \begin{axis}[
            xlabel={Number of Nodes},
            ylabel={Throughput},
            legend style={at={(0.5,-0.25)}, 
            anchor=north,legend columns=-1},
            symbolic x coords={500, 1000, 2000, 3000, 4000, 5000},
            xtick=data,
            mark size=2, 
            mark options={solid}, 
        ]
        \addplot[dashed, mark=*] coordinates {(500, 318) (1000, 317) (2000, 318) (3000, 317) (4000, 318) (5000, 317)};
        \addplot[dotted, mark=triangle] coordinates {(500, 342) (1000, 342) (2000, 342) (3000, 342) (4000, 341) (5000, 342)};
        \addplot[dashdotted, mark=square] coordinates {(500, 340) (1000, 340) (2000, 340) (3000, 340) (4000, 340) (5000, 341)};
        \addplot[solid, mark=diamond] coordinates {(500, 342) (1000, 341) (2000, 341) (3000, 342) (4000, 342) (5000, 342)};
        \legend{PAXOS, RAFT, PBFT, BECP}
        \end{axis}
    \end{tikzpicture}
    \caption{Comparative Throughput Analysis between Traditional Protocols (PAXOS, RAFT, PBFT) and BECP: Impact of Node Scalability (N = 500 to 5000) over a Simulation Time of 3600 Seconds}
    \label{fig:throughput_1}
\end{figure}
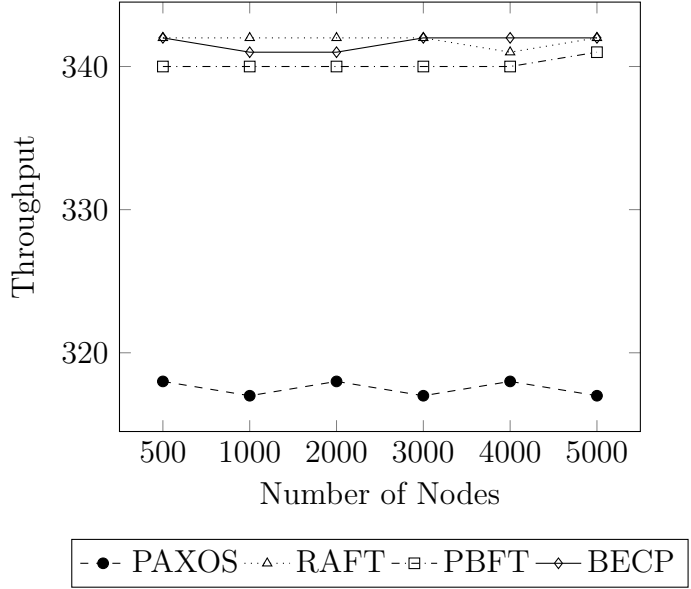
\begin{figure}
    \centering
    \begin{tikzpicture}
        \begin{axis}[ 
            xlabel={Number of Nodes},
            ylabel={Throughput},
            legend style={at={(0.5,-0.25)}, 
            anchor=north,legend columns=-1},
            symbolic x coords={500, 1000, 2000, 3000, 4000, 5000},
            xtick=data,
            mark size=2, 
            mark options={solid}, 
        ]
        \addplot[solid, mark=diamond] coordinates {(500, 56) (1000, 56) (2000, 56) (3000, 56) (4000, 55) (5000, 55)};
        \addplot[dashdotted, mark=square] coordinates {(500, 48) (1000, 47) (2000, 47) (3000, 46) (4000, 47) (5000, 45)};
        \legend{BECP, Avalanche}
        \end{axis}
    \end{tikzpicture}
    \caption{Comparative Throughput Analysis between BECP and Avalanche: Impact of Node Scalability (N = 500 to 5000) over a Simulation Time of 600 Seconds (blocks for BECP, transactions for Avalanche)}
    \label{fig:throughput_2}
\end{figure}
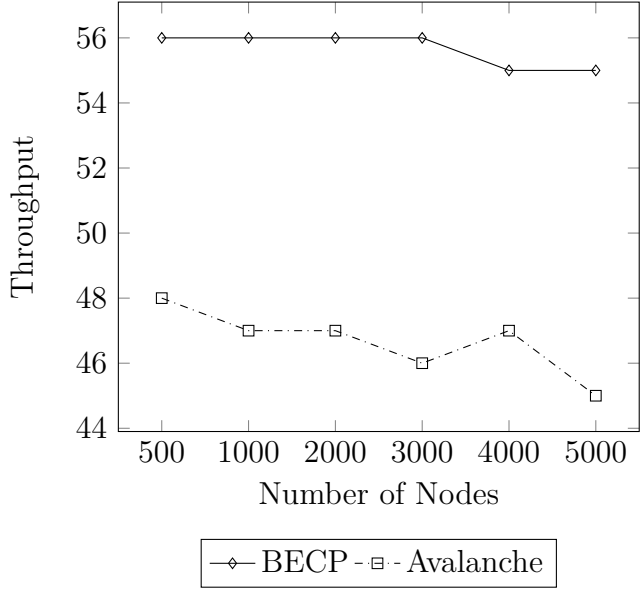
Figure \ref{fig:throughput_1} depicts the throughput comparison for system sizes ranging from $500$ to $5000$ nodes among the traditional protocols and BECP. RAFT, PBFT, and BECP exhibit the highest throughput, highlighting the efficiency of these protocols. Figure \ref{fig:throughput_2} presents a similar comparison between Avalanche and BECP, showing that BECP achieves higher throughput. Considering that Avalanche utilizes transactions instead of blocks, and each block can contain nearly $2000$ transactions, BECP demonstrates greater efficiency compared to Avalanche.

\textbf{Communication Overhead:} Communication Overhead refers to the traffic of the network or the total number of sent messages by nodes in a given amount of time. This metric is important since network congestion increases the delay of sent messages, including block propagation, and as a result, reduces the efficiency of the protocol. Figure \ref{fig:overhead_1} and Figure \ref{fig:overhead_2} illustrate the comparison of the communication overhead metric of the protocols. 
\begin{figure}
    \centering
    \begin{tikzpicture}
        \begin{axis}[
            xlabel={Number of Nodes},
            ylabel={Number of Sent Messages},
            legend style={at={(0.5,-0.25)}, 
            anchor=north,legend columns=-1},
            symbolic x coords={500, 1000, 2000, 3000, 4000, 5000},
            xtick=data,
            mark size=2, 
            mark options={solid}, 
            ymode=log,
        ]
        \addplot[dashed, mark=*] coordinates {(500, 795273) (1000, 1589543) (2000, 3178770) (3000, 4768290) (4000, 6358447) (5000, 7947432)};
        \addplot[dotted, mark=triangle] coordinates {(500, 3330671) (1000, 7161382) (2000, 16322820) (3000, 27484462) (4000, 40645754) (5000, 55807159)};
        \addplot[dashdotted, mark=square] coordinates {(500, 171375495) (1000, 685424970) (2000, 2740358265) (3000, 6166159545) (4000, 10959916245) (5000, 17123732775)};
        \addplot[solid, mark=diamond] coordinates {(500, 5135750) (1000, 10271503) (2000, 20542982) (3000, 30814476) (4000, 41085959) (5000, 51357431)};
        \legend{PAXOS, RAFT, PBFT, BECP}
        \end{axis}
    \end{tikzpicture}
    \caption{Comparative Communication Overhead Analysis between Traditional Protocols (PAXOS, RAFT, PBFT) and BECP: Impact of Node Scalability (N = 500 to 5000) over a Simulation Time of 3600 Seconds}
    \label{fig:overhead_1}
\end{figure}
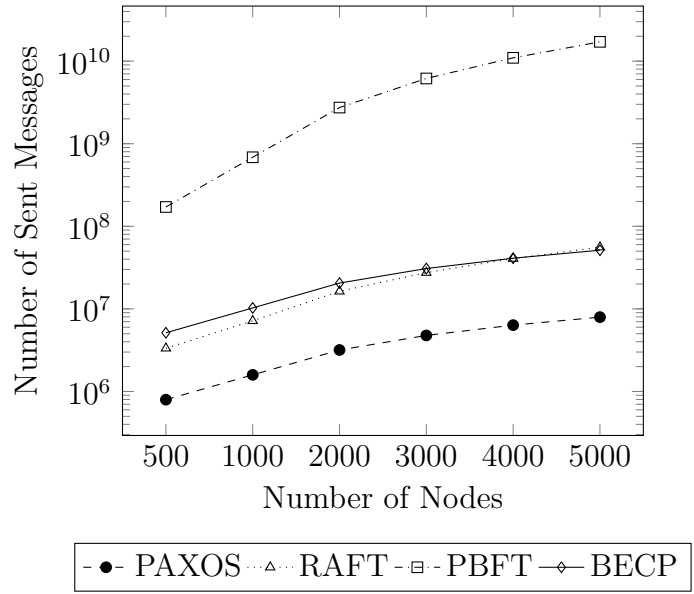
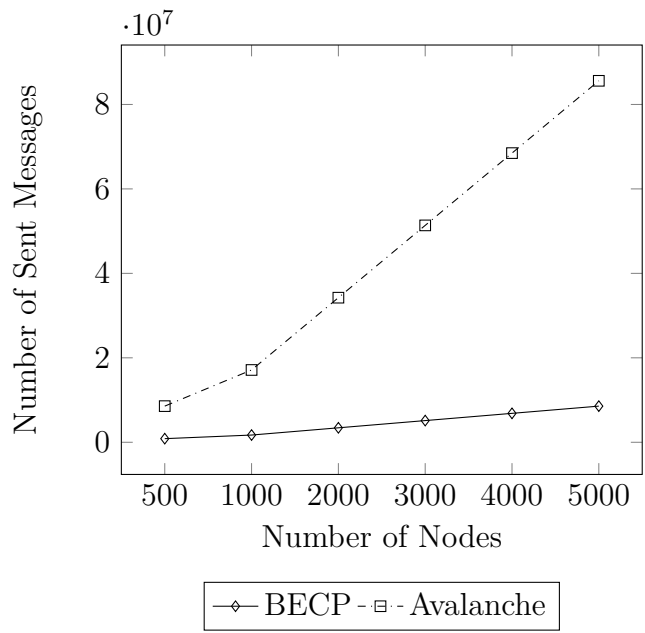
\begin{figure}
    \centering
    \begin{tikzpicture}
        \begin{axis}[
            xlabel={Number of Nodes},
            ylabel={Number of Sent Messages},
            legend style={at={(0.5,-0.25)}, 
            anchor=north,legend columns=-1},
            symbolic x coords={500, 1000, 2000, 3000, 4000, 5000},
            xtick=data,
            mark size=2, 
            mark options={solid}, 
        ]
        \addplot[solid, mark=diamond] coordinates {(500, 855999) (1000, 1711998) (2000, 3423995) (3000, 5135990) (4000, 6847990) (5000, 8559984)};
        \addplot[dashdotted, mark=square] coordinates {(500, 8559988) (1000, 17119967) (2000, 34239933) (3000, 51359899) (4000, 68479872) (5000, 85599845)};
        \legend{BECP, Avalanche}
        \end{axis}
    \end{tikzpicture}
    \caption{Comparative Communication Overhead Analysis between BECP and Avalanche: Impact of Node Scalability (N = 500 to 5000) over a Simulation Time of 600 Seconds}
    \label{fig:overhead_2}
\end{figure}
According to Figure \ref{fig:overhead_1}, PBFT demonstrates the highest communication overhead, while PAXOS exhibits the lowest. BECP and RAFT show nearly the same level of communication overhead. However, a comparison between BECP and Avalanche reveals that BECP has lower communication overhead, as shown in Figure \ref{fig:overhead_2}. For all protocols, communication overhead increases as the number of nodes (system size) increases. However, for PBFT and Avalanche, this increase is more significant.

\textbf{Consensus Latency:} 
Consensus latency is defined as the time from block generation to acceptance. A protocol with lower consensus latency is considered a desirable protocol. In the experiments, we considered an average of consensus latency for the confirmed blocks. Figure \ref{fig:latency_1} and Figure \ref{fig:latency_2} illustrate the comparison of the average consensus latency metric among the studied protocols. As evident in the figures, PAXOS, RAFT, and PBFT exhibit very small latency compared to BECP and Avalanche. However, BECP demonstrates lower latency compared to Avalanche.
\begin{figure}
    \centering
    \begin{tikzpicture}
        \begin{axis}[
            xlabel={Number of Nodes},
            ylabel={Average Consensus Latency (seconds)},
            legend style={at={(0.5,-0.25)}, 
            anchor=north,legend columns=-1},
            symbolic x coords={500, 1000, 2000, 3000, 4000, 5000},
            xtick=data,
            mark size=2, 
            mark options={solid}, 
        ]
        \addplot[dashed, mark=*] coordinates {(500, 0.738) (1000, 0.743) (2000, 0.746) (3000, 0.748) (4000, 0.749) (5000, 0.749)};
        \addplot[dotted, mark=triangle] coordinates {(500, 0.467) (1000, 0.458) (2000, 0.455) (3000, 0.448) (4000, 0.436) (5000, 0.445)};
        \addplot[dashdotted, mark=square] coordinates {(500, 0.566) (1000, 0.566) (2000, 0.566) (3000, 0.567) (4000, 0.566) (5000, 0.566)};
        \addplot[solid, mark=diamond] coordinates {(500, 20.306) (1000, 20.821) (2000, 21.281) (3000, 21.526) (4000, 21.7) (5000, 21.839)};
        \legend{PAXOS, RAFT, PBFT, BECP}
        \end{axis}
    \end{tikzpicture}
    \caption{Comparative Consensus Latency Analysis between Traditional Protocols (PAXOS, RAFT, PBFT) and BECP: Impact of Node Scalability (N = 500 to 5000) over a Simulation Time of 3600 Seconds}
    \label{fig:latency_1}
\end{figure}
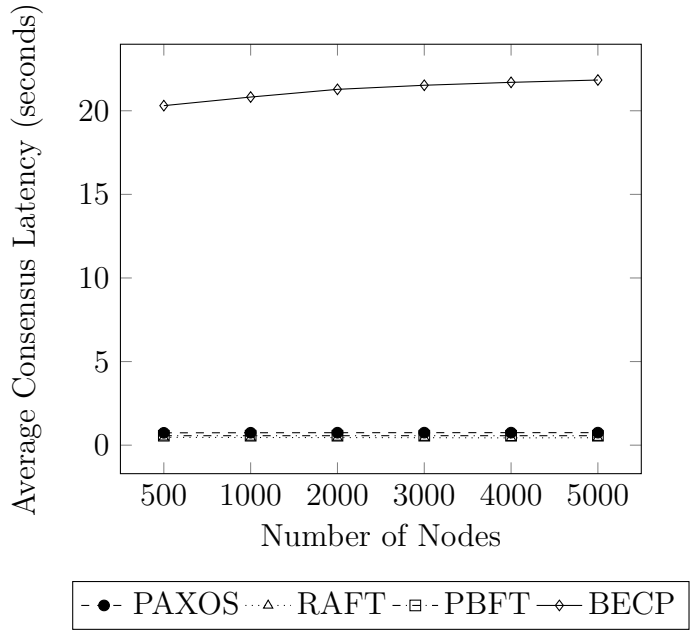
\begin{figure}
    \centering
    \begin{tikzpicture}
        \begin{axis}[ 
            xlabel={Number of Nodes},
            ylabel={Average Consensus Latency (seconds)},
            legend style={at={(0.5,-0.25)}, 
            anchor=north,legend columns=-1},
            symbolic x coords={500, 1000, 2000, 3000, 4000, 5000},
            xtick=data,
            mark size=2, 
            mark options={solid}, 
        ]
        \addplot[solid, mark=diamond] coordinates {(500, 20.286) (1000, 20.798) (2000, 21.249) (3000, 21.522) (4000, 21.719) (5000, 21.812)};
        \addplot[dashdotted, mark=square] coordinates {(500, 101.915) (1000, 101.212) (2000, 102.968) (3000, 98.831) (4000, 99.457) (5000, 103.99)};
        \legend{BECP, Avalanche}
        \end{axis}
    \end{tikzpicture}
    \caption{Comparative Consensus Latency Analysis between BECP and Avalanche: Impact of Node Scalability (N = 500 to 5000) over a Simulation Time of 600 Seconds}
    \label{fig:latency_2}
\end{figure}
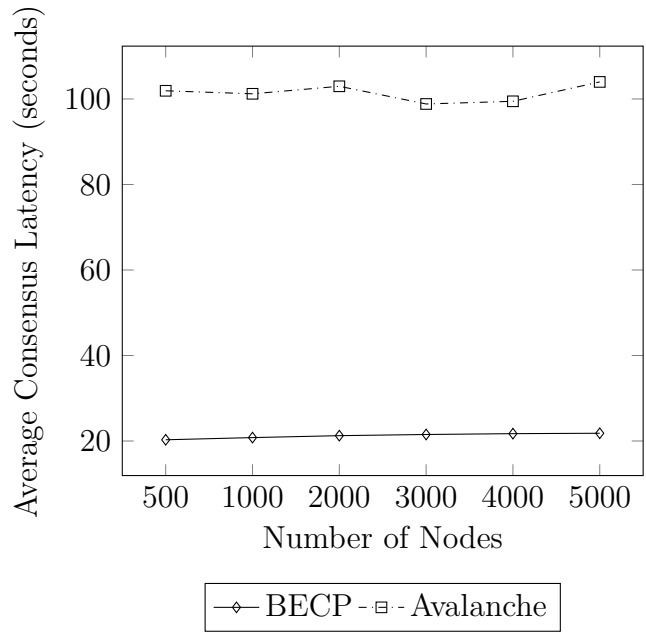
\section{Discussion}
\label{sec:discussion}
In this section, we provide a comprehensive discussion of the presented results. According to Figure \ref{fig:throughput_1}, protocols RAFT, PBFT, and BECP achieved a nearly similar throughput rate, while the protocol PAXOS achieved a slightly lower rate. This disparity can be attributed to the structure of PAXOS, where the promise phase is performed at every round of the consensus process. This frequent promise phase brings latency into the process, resulting in a lower throughput rate. In contrast, algorithm RAFT mitigates this issue by selecting a leader only once, leading to a higher throughput rate.

Figure \ref{fig:throughput_2} demonstrates the efficiency of BECP over Avalanche. BECP achieves a slightly higher throughput rate due to its consensus mechanism, which leverages an epidemic diffusion process and local estimations for quicker consensus. During each cycle, nodes simultaneously send and receive information, further boosting performance.

When comparing the communication overhead of the protocols (Figures \ref{fig:overhead_1} and \ref{fig:overhead_2}), we found that PBFT exhibits the highest overhead among the others due to its communication complexity. This overhead primarily stems from its communication complexity, which necessitates a broadcast of votes from all nodes at every phase of the protocol. 

Furthermore, protocol RAFT incurs a higher overhead compared to PAXOS. This is because in RAFT, in addition to communication messages for the consensus process, the leader node also broadcasts regular heartbeat messages in the network to maintain its leadership. Overall, PAXOS and RAFT protocols exhibit lower communication overhead due to their centralized structure, where only the leader node sends the blocks. In contrast, protocols BECP and Avalanche demonstrate higher communication overhead compared to PAXOS and RAFT. In these protocols, nodes can generate new blocks with a probability without waiting for the confirmation of previous ones. On the other hand, by comparing the communication overhead of BECP and Avalanche, it is evident that BECP has a lower overhead compared to Avalanche because, in Avalanche, each node draws samples from $K$ neighbours. In contrast, nodes in BECP send a message to only one neighbour.

The traditional protocols exhibit very low average consensus latency (Figure \ref{fig:latency_1}) due to their deterministic structure. They can achieve consensus on an item in three message passes, apart from the leader election process. However, this approach involves achieving consensus on blocks one after another. BECP and Avalanche demonstrate higher average consensus latency (Figure \ref{fig:latency_2}) because, unlike traditional protocols, they do not rely on the direct communication assumption. This approach closely resembles real-world applications, where users can generate blocks with a probability. However, as depicted in Figure \ref{fig:latency_2}, BECP has lower latency compared to Avalanche, while achieving the same throughput rate as the traditional protocols. This efficiency is due to its epidemic consensus mechanism, which broadcasts blocks throughout the network, resulting in faster consensus.

Based on the results obtained from $500$ to $5000$ nodes, we found that the protocol BECP demonstrates desirable scalability properties compared to the other tested protocols. BECP maintains a steady rate of throughput, which surpasses Avalanche's rate of throughput. Additionally, it exhibits lower communication overhead compared to PBFT and Avalanche.
\section{Conclusion}
\label{sec:conclusion}
In this paper, the Blockchain Epidemic Consensus Protocol (BECP) protocol has been introduced. BECP is a novel and fully decentralised consensus system for blockchain networks, that leverages epidemic communication and local computation to facilitate consensus on blocks. Unlike traditional protocols such as PAXOS, RAFT, and PBFT, BECP operates without a designated leader, thereby ensuring robust decentralisation. Unlike PoW mechanisms, BECP imposes minimal resource demands, and unlike PoS, it is not vulnerable to collusion. Our simulations and analyses confirm that BECP shows favourable metrics including throughput, scalability, communication overhead, and consensus latency when compared to existing protocols. Future efforts will focus on augmenting BECP consensus mechanism to include the ability to detect node failures and trigger a recovery mechanism to provide system resilience towards these failures.
\bibliographystyle{plain}
\bibliography{main.bib}
\end{document}